\documentclass{article}

\usepackage{spconf,amsmath,graphicx,bm}
\usepackage{booktabs}

\usepackage{caption}
\title{MISPRONUNCIATION DETECTION IN NON-NATIVE (L2) ENGLISH WITH UNCERTAINTY MODELING}
\name{\begin{tabular}{c}Daniel Korzekwa$^{\star \dagger}$, Jaime Lorenzo-Trueba$^{\star}$, Szymon Zaporowski$^{\dagger}$, \\
Shira Calamaro$^{\star}$, Thomas Drugman$^{\star}$, Bozena Kostek$^{\dagger}$\end{tabular}}
\address{$^{\star}$ Amazon Speech $^{\dagger}$ Gdansk University of Technology, Faculty of ETI, Poland}
\begin{document}
%
\maketitle
\begin{abstract}
A common approach to the automatic detection of mispronunciation in language learning is to recognize the phonemes produced by a student and compare it to the expected pronunciation of a native speaker. This approach makes two simplifying assumptions:  a) phonemes can be recognized from speech with high accuracy, b) there is a single correct way for a sentence to be pronounced. These assumptions do not always hold, which can result in a significant amount of false mispronunciation alarms. We propose a novel approach to overcome this problem based on two principles: a) taking into account uncertainty in the automatic phoneme recognition step, b) accounting for the fact that there may be multiple valid pronunciations. We evaluate the model on non-native (L2) English speech of German, Italian and Polish speakers, where it is shown to increase the precision of detecting mispronunciations by up to 18\% (relative) compared to the common approach.
\end{abstract}
\section{Introduction}
\label{sec:intro}

In Computer Assisted Pronunciation Training (CAPT), students are presented with a text and asked to read it aloud. A computer informs students on mispronunciations in their speech, so that they can repeat it and improve. CAPT has been found to be an effective tool that helps non-native (L2) speakers of English to improve their pronunciation skills \cite{neri2008effectiveness,tejedor2020assessing}.

A common approach to CAPT is based on recognizing the phonemes produced by a student and comparing them with the expected (canonical) phonemes that a native speaker would pronounce \cite{witt2000phone,li2016mispronunciation,sudhakara2019improved,leung2019cnn}.  It makes two simplifying assumptions. First, it assumes that phonemes can be automatically recognized from speech with high accuracy. However, even in native (L1) speech, it is difficult to get the Phoneme Error Rate (PER) below 15\% \cite{chorowski2015attention}. Second, this approach assumes that this is the only `correct' way for a sentence to be pronounced, but due to phonetic variability this is not always true. For example, the word `enough' can be pronounced by native speakers in multiple correct ways:  /ih n ah f/ or /ax n ah f/ (short `i' or `schwa' phoneme at the beginning). These assumptions do not always hold which can result in a significant amount of false mispronunciation alarms and making students confused when it happens.

We propose a novel approach that results in fewer false mispronunciation alarms, by formalizing the intuition that we will not be able to recognize exactly what a student has pronounced or say precisely how a native speaker would pronounce it. First, the model estimates a belief over the phonemes produced by the student, intuitively representing the uncertainty in the student's pronunciation. Then, the model converts this belief into the probabilities that a native speaker would pronounce it, accounting for phonetic variability. Finally, the model makes a decision on which words were mispronounced in the sentence by processing three pieces of information: a) what the student pronounced, b) how likely a native speaker would pronounce it that way, and c) what the student was expected to pronounce. 

In Section \ref{sec:related_work}, we review the related work. In Section \ref{sec:proposed_model}, we describe the proposed model. In Section \ref{sec:experiments}, we present the experiments, and we conclude in Section \ref{sec:conclusion}.

\section{Related Work}
\label{sec:related_work}

In 2000, Witt et al. coined the term Goodness of Pronunciation (GoP) \cite{witt2000phone}. GoP starts by aligning the canonical phonemes with the speech signal using a forced-alignment technique. This technique aims to find the most likely mapping between phonemes and the regions of a corresponding speech signal. In the next step, GoP computes the ratio between the likelihoods of the canonical and the most likely pronounced phonemes. Finally, it detects a mispronunciation if the ratio falls below a given threshold. GoP was further extended with Deep Neural Networks (DNNs), replacing Hidden Markov Model (HMM) and Gaussian Mixture Model (GMM) techniques for acoustic modeling \cite{li2016mispronunciation,sudhakara2019improved}. Cheng et al. \cite{cheng2020asr} improved the performance of GoP with the latent representation of speech extracted in an unsupervised way.

As opposed to GoP, we do not use forced-alignment that requires both speech and phoneme inputs. Following the work of Leung et al. \cite{leung2019cnn}, we use a phoneme recognizer, which recognizes phonemes from only the speech signal. The phoneme recognizer is based on a Convolutional Neural Network (CNN), a Gated Recurrent Unit (GRU), and Connectionist Temporal Classification (CTC) loss. Leung et al. report that it outperforms other forced-alignment \cite{li2016mispronunciation} and forced-alignment-free \cite{harrison2009implementation} techniques on the task of detecting phoneme-level mispronunciations in L2 English. Contrary to Leung et al., who rely only on a single recognized sequence of phonemes, we obtain top $N$ decoded sequences of phonemes, along with the phoneme-level posterior probabilities.

It is common in pronunciation assessment to employ the speech signal of a reference speaker. Xiao et al. use a pair of speech signals from a student and a native speaker to classify native and non-native speech \cite{xiao2018paired}. Mauro et al. incorporate the speech of a reference speaker to detect mispronunciations at the phoneme level \cite{nicolao2015automatic}. Wang et al. use siamese networks for modeling discrepancy between normal and distorted children's speech \cite{wang2019child}. We take a similar approach but we do not need a database of reference speech. Instead, we train a statistical model to estimate the probability of pronouncing a sentence by a native speaker. Qian et al. propose a statistical pronunciation model as well \cite{qian2010capturing}. Unlike our work, in which we create a model of `correct` pronunciation, they build a model that generates hypotheses of mispronounced speech.

\section{Proposed Model}
\label{sec:proposed_model}

\begin{figure*}[htb]
  \centering
  \includegraphics[height=2.6cm]{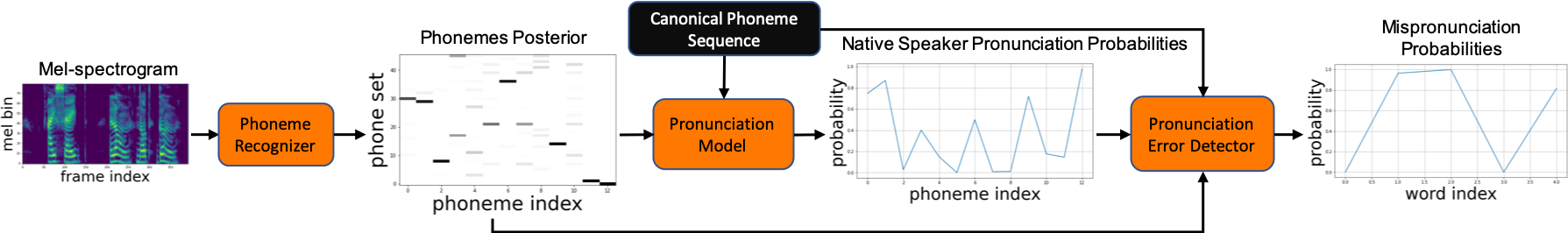}
  \caption{Architecture of the system for detecting mispronounced words in a spoken sentence.}
  \label{fig:model_architecture}
\end{figure*}

The design consists of three subsystems: a Phoneme Recognizer (PR), a Pronunciation Model (PM), and a Pronunciation Error Detector (PED), illustrated in Figure \ref{fig:model_architecture}. The PR recognizes phonemes spoken by a student. The PM estimates the probabilities of having been pronounced by a native speaker. Finally, the PED computes word-level mispronunciation probabilities. In Figure \ref{fig:nn_architecture}, we present detailed architectures of the PR, PM, and PED.

For example, considering the text: `I said alone not gone' with the canonical representation of /ay - s eh d - ax l ow n - n aa t - g aa n/. Polish L2 speakers of English often mispronounce the /eh/ phoneme in the second word as /ey/. The PM would identify the /ey/ as having a low probability of being pronounced by a native speaker in the middle of the word `said’, which the PED would translate into a high probability of mispronunciation.

\subsection{Phoneme Recognizer}
\label{ssec:phoneme_recognizer}

\begin{figure*}[htb]
  \centering
  \includegraphics[height=4.0cm]{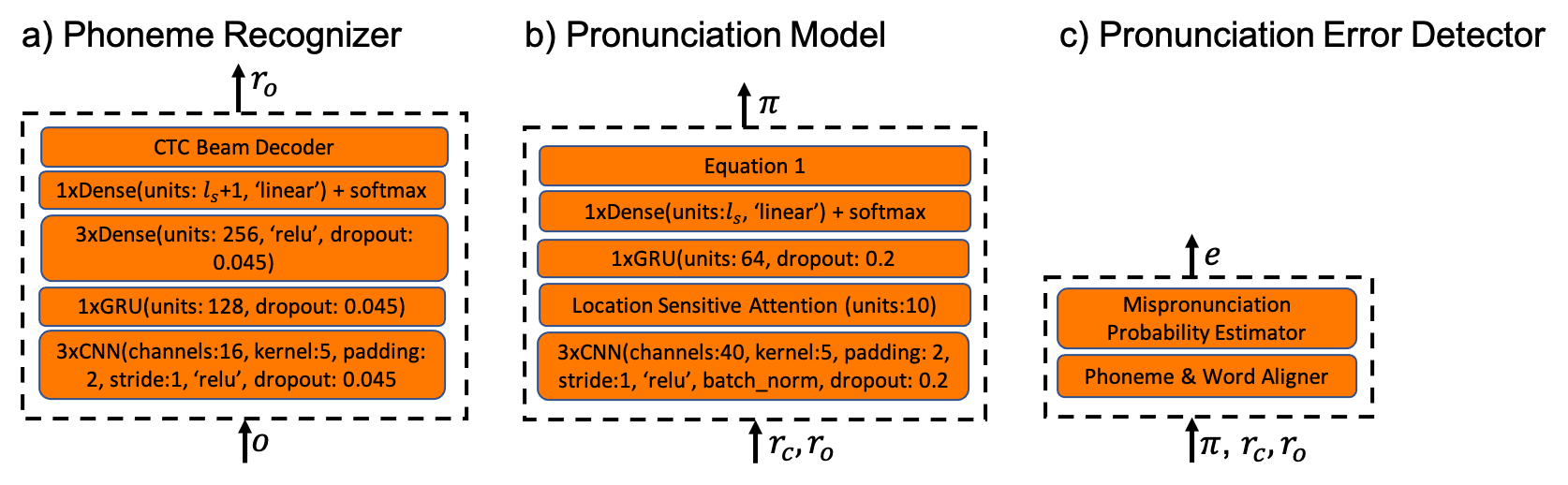}
  \caption{Architecture of the PR, PM, and PED subsystems. $l_s$ - the size of the phoneme set.}
  \label{fig:nn_architecture}
\end{figure*}

The PR (Figure \ref{fig:nn_architecture}a) uses beam decoding \cite{graves2013speech} to estimate $N$ hypotheses of the most likely sequences of phonemes that are recognized in the speech signal $\mathbf{o}$. A single hypothesis is denoted as $\mathbf{r_o} \sim p(\mathbf{r_o}|\mathbf{o})$. The speech signal $\mathbf{o}$ is represented by a mel-spectrogram with $f$ frames and 80 mel-bins. Each sequence of phonemes $\mathbf{r_o}$ is accompanied by the posterior phoneme probabilities of shape: $(l_{r_o},l_s+1)$. $l_{r_o}$ is the length of the sequence and $l_s$ is the size of the phoneme set (45 phonemes including `pause', `end of sentence (eos)', and a `blank' label required by the CTC-based model).

\subsection{Pronunciation Model}
\label{ssec:proonunciation_model}

The PM (Figure \ref{fig:nn_architecture}b) is an encoder-decoder neural network following Sutskever et al. \cite{sutskever2014sequence}. Instead of building a text-to-text translation system between two languages, we use it for phoneme-to-phoneme conversion. The sequence of phonemes $\mathbf{r_c}$ that a native speaker was expected to pronounce is converted into the sequence of phonemes  $\mathbf{r}$ they had pronounced, denoted as  $\mathbf{r} \sim p(\mathbf{r}|\mathbf{r_c})$. Once trained, the PM acts as a probability mass function, computing the likelihood sequence $\bm{\pi}$ of the phonemes $\mathbf{r_o}$ pronounced by a student conditioned on the expected (canonical) phonemes $\mathbf{r_c}$. The PM is denoted in Eq. \ref{eq_proposed_model}, which we implemented in MxNet \cite{chen2015mxnet} using `sum' and `element-wise multiply' linear-algebra operations.

\begin{equation}
\bm{\pi}=\sum_{\mathbf{r_o}} p(\mathbf{r_o}|\mathbf{o})p(\mathbf{r}=\mathbf{r_o}|\mathbf{r_c})
\label{eq_proposed_model}
\end{equation}

The model is trained on phoneme-to-phoneme speech data created automatically by passing the speech of the native speakers through the PR. By annotating the data with the PR, we can make the PM model more resistant to possible phoneme recognition inaccuracies of the PR at testing time.

\subsection{Pronunciation Error Detector}
\label{ssec:pron_error_detector}

The PED (Figure \ref{fig:nn_architecture}c) computes the probabilities of mispronunciations $\mathbf{e}$ at the word level, denoted as $\mathbf{e} \sim p(\mathbf{e}|\mathbf{r_o},\bm{\pi},\mathbf{r_c})$. The PED is conditioned on three inputs: the phonemes $\mathbf{r_o}$ recognized by the PR, the corresponding pronunciation likelihoods $\bm{\pi}$ from the PM, and the canonical phonemes $\mathbf{r_c}$. The model starts with aligning the canonical and recognized sequences of phonemes. We adopted a dynamic programming algorithm for aligning biological sequences developed by Needleman-Wunsch \cite{needleman1970general}. Then, the probability of mispronunciation for a given word is computed with Equation \ref{eq_ped}, $k$ denotes the word index, and $j$ is the phoneme index in the word with the lowest probability of pronunciation.

\begin{equation}
p(\mathbf{e}_k) =
    \begin{cases}
      0 & \text{if aligned phonemes match,}\\
       1-\bm{\pi}_{k,j}  & \text{otherwise.}
    \end{cases} 
 \label{eq_ped}
\end{equation}

We compute the probabilities of mispronunciation for $N$ phoneme recognition hypotheses from the PR. Mispronunciation for a given word is detected if the probability of mispronunciation falls below a given threshold for all hypotheses. The hyper-parameter $N=4$ was manually tuned on a single L2 speaker from the testing set to optimize the PED in the precision metric.
 
\section{Experiments and Discussion}
\label{sec:experiments}

We want to understand the effect of accounting for uncertainty in the PR-PM system presented in Section \ref{sec:proposed_model}. To do this, we compare it with two other variants, PR-LIK and PR-NOLIK, and analyze precision and recall metrics. The PR-LIK system helps us understand how important is it to account for the phonetic variability in the PM. To switch the PM off, we modify it so that it considers only a single way for a sentence to be pronounced correctly. 

The PR-NOLIK variant corresponds to the CTC-based mispronunciation detection model proposed by Leung et al. \cite{leung2019cnn}. To reflect this, we make two modifications compared to the PR-PM system. First, we switch the PM off in the same way we did it in the PR-LIK system. Second, we set the posterior probabilities of recognized phonemes in the PR to 100\%, which means that the PR is always certain about the phonemes produced by a speaker. There are some slight implementation differences between Leung's model and PR-NOLIK, for example, regarding the number of units in the neural network layers. We use our configuration to make a consistent comparison with PR-PM and PR-LIK systems. One can hence consider PR-NOLIK as a fair state-of-the-art baseline \cite{leung2019cnn}.

\subsection{Model Details}
\label{ssec:model_details}

For extracting mel-spectrograms, we used a time step of 10 ms and a window size of 40 ms. The PR was trained with CTC Loss and Adam Optimizer (batch size: 32, learning rate: 0.001, gradient clipping: 5). We tuned the following hyper-parameters of the PR with Bayesian Optimization: dropout, CNN channels, GRU, and dense units. The PM was trained with the cross-entropy loss and AdaDelta optimizer (batch size: 20, learning rate: 0.01, gradient clipping: 5). The location-sensitive attention in the PM follows the work by Chorowski et al. \cite{chorowski2015attention}. The PR and PM models were implemented in MxNet Deep Learning framework.

\subsection{Speech Corpora}
\label{ssec:speech_corpora}

For training and testing the PR and PM, we used 125.28 hours of L1 and L2 English speech from 983 speakers segmented into 102812 sentences, sourced from multiple speech corpora: TIMIT \cite{garofolo1993darpa}, LibriTTS \cite{Zen2019}, Isle \cite{atwell2003isle} and GUT Isle \cite{Weber2020}. We summarize it in Table \ref{tab:speech_corpora}. All speech data were downsampled to 16 kHz. Both L1 and L2 speech were phonetically transcribed using Amazon proprietary grapheme-to-phoneme model and used by the PR. Automatic transcriptions of L2 speech do not capture pronunciation errors, but we found it is still worth including automatically transcribed L2 speech in the PR. L2 corpora were also annotated by 5 native speakers of American English for word-level pronunciation errors. There are 3624 mispronounced words out of 13191 in the Isle Corpus and 1046 mispronounced words out of 5064 in the GUT Isle Corpus.

From the collected speech, we held out 28 L2 speakers and used them only to assess the performance of the systems in the mispronunciation detection task. It includes 11 Italian and 11 German speakers from the Isle corpus \cite{atwell2003isle}, and 6 Polish speakers from the GUT Isle corpus \cite{Weber2020}.

\begin{table}[htb]
\scriptsize
\caption{The summary of speech corpora used by the PR.}
  \label{tab:speech_corpora}
  \centering
  \begin{tabular}{lll}
    \toprule  
   Native Language & Hours & Speakers \\
    \midrule
    English & 90.47 & 640\\ 
    Unknown & 19.91 & 285\\  
    German and Italian & 13.41 & 46\\  
    Polish & 1.49 & 12\\ 
    \bottomrule
  \end{tabular}
\end{table}

\subsection{Experimental Results}
\label{ssec:experiment_results}

The PR-NOLIK detects mispronounced words based on the difference between the canonical and recognized phonemes. Therefore, this system does not offer any flexibility in optimizing the model for higher precision. 

\begin{figure}[!tb]
  \centering
  \includegraphics[height=2.7cm]{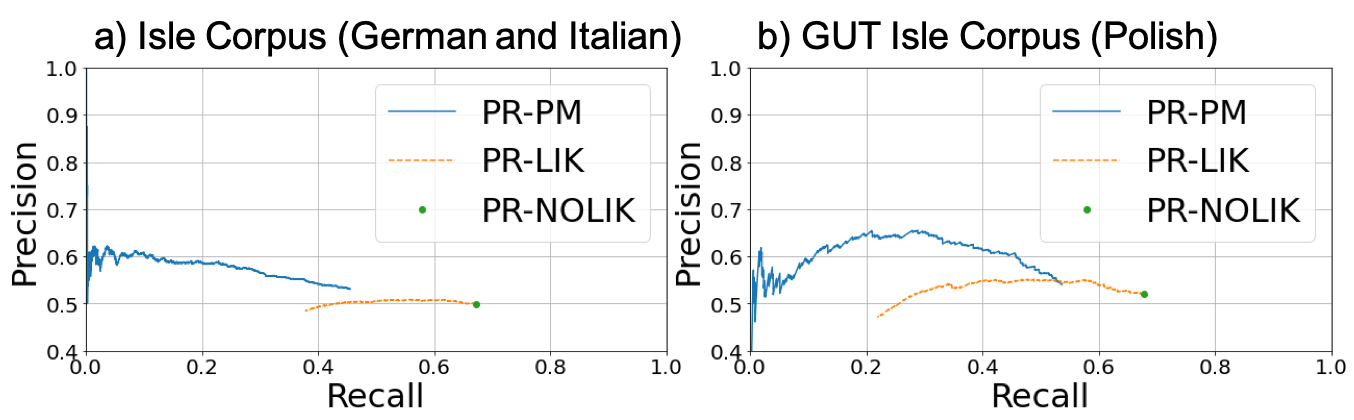}
  \caption{Precision-recall curves for the evaluated systems.}
  \label{fig:precision_recall_plot}
\end{figure}

The PR-LIK system incorporates posterior probabilities of recognized phonemes. It means that we can tune this system towards higher precision, as illustrated in Figure \ref{fig:precision_recall_plot}. Accounting for uncertainty in the PR helps when there is more than one likely sequence of phonemes that could have been uttered by a user, and the PR model is uncertain which one it is. For example, the PR reports two likely pronunciations for the text `I said' /ay s eh d/. The first one, /s eh d/ with /ay/ phoneme missing at the beginning and the alternative one  /ay s eh d/ with the /ay/ phoneme present. If the PR considered only the mostly likely sequence of phonemes, like PR-NOLIK does, it would incorrectly raise a pronunciation error. In the second example, a student read the text `six' /s ih k s/ mispronouncing the first phoneme /s/ as /t/. The likelihood of the recognized phoneme is only 34\%. It suggests that the PR model is quite uncertain on what phoneme was pronounced. However, sometimes even in such cases, we can be confident that the word was mispronounced. It is because the PM computes the probability of pronunciation based on the posterior probability from the PR model. In this particular case, other phoneme candidates that account for the remaining 66\% of uncertainty are also unlikely to be pronounced by a native speaker. The PM can take it into account and correctly detect a mispronunciation.

However, we found that the effect of accounting for uncertainty in the PR is quite limited. Compared to the PR-NOLIK system, the PR-LIK raises precision on the GUT Isle corpus only by 6\% (55\% divided by 52\%), at the cost of dropping recall by about 23\%. We can observe a much stronger effect when we account for uncertainty in the PM model. Compared to the PR-LIK system, the PR-PM system further increases precision between 11\% and 18\%, depending on the decrease in recall between 20\% to 40\%. One example where the PM helps is illustrated by the word `enough' that can be pronounced in two similar ways: /ih n ah f/ or /ax n ah f/ (short `i' or `schwa' phoneme at the beginning.) The PM can account for phonetic variability and recognize both versions as pronounced correctly. Another example is word linking \cite{hieke1984linking}. Native speakers tend to merge phonemes of neighboring words. For example, in the text `her arrange' /hh er - er ey n jh/, two neighboring phonemes /er/ can be pronounced as a single phoneme: /hh er ey n jh/. The PM model can correctly recognize multiple variations of such pronunciations.

Complementary to precision-recall curve showed in Figure \ref{fig:precision_recall_plot}, we present in Table \ref{tab:pron_error_detection} one configuration of the precision and recall scores for the PR-LIK and PR-PM systems. This configuration is selected in such a way that:  a) recall for both systems is close to the same value, b) to illustrate that the PR-PM model has a much bigger potential of increasing precision than the PR-LIK system. A similar conclusion can be made by inspecting multiple different precision and recall configurations in the precision and recall plots for both Isle and GUT Isle corpora.

\begin{table}[thb]
\scriptsize
\scriptsize
\caption{Precision and recall of detecting word-level mispronunciations. CI - Confidence Interval.}
  \label{tab:pron_error_detection}
  \centering
  \begin{tabular}{lll}
    \toprule  
   Model  & Precision [\%,95\%CI] & Recall [\%,95\%CI] \\
    \midrule
    \multicolumn{3}{c}{\textbf{Isle corpus (German and Italian)}} \\
    PR-LIK & 49.39 (47.59-51.19) & 40.20 (38.62-41.81)\\ 
    PR-PM & 54.20 (52.32-56.08) & 40.20 (38.62-41.81)\\  
  
     \multicolumn{3}{c}{\textbf{GUT Isle corpus (Polish)}} \\
    PR-LIK & 54.91 (50.53-59.24) & 40.29 (36.66-44.02)\\ 
    PR-PM & 61.21 (56.63-65.65) & 40.15 (36.51-43.87)\\  
    
    \bottomrule
  \end{tabular}
\end{table}
 
\section{Conclusion and Future Work}
\label{sec:conclusion}

To report fewer false pronunciation alarms, it is important to move away from the two simplifying assumptions that are usually made by common methods for pronunciation assessment: a) phonemes can be recognized with high accuracy, b) a sentence can be read in a single correct way. We acknowledged that these assumptions do not always hold. Instead, we designed a model that: a) accounts for the uncertainty in phoneme recognition and b) accounts for multiple ways a sentence can be pronounced correctly due to phonetic variability. We found that to optimize precision, it is more important to account for the phonetic variability of speech than accounting for uncertainty in phoneme recognition. We showed that the proposed model can raise the precision of detecting mispronounced words by up to 18\% compared to the common methods.

In the future, we plan to adapt the PM model to correctly pronounced L2 speech to account for phonetic variability of non-native speakers. We plan to combine the PR, PM, and PED modules and train the model jointly to eliminate accumulation of statistical errors coming from disjoint training of the system.

\bibliographystyle{IEEEbib}
\bibliography{strings,refs,pronunciation_uncertainty_modeling}

\end{document}